\documentclass[twoside]{proc}

\usepackage{algorithm}
\usepackage{algpseudocode}
\usepackage{amsmath}
\usepackage{amssymb}
\usepackage{amsmath,epsfig,bm,subfigure}
\usepackage{multirow}
\usepackage{tabu}
\usepackage{makecell}
\usepackage{epstopdf}
\usepackage{cite}
\usepackage{threeparttable}
\usepackage{url}
\usepackage{booktabs,bigstrut}

\begin{document}
\title{Learned Video Compression with Residual Prediction and Loop Filter}
\author{Chao~Liu,
        Heming~Sun\thanks{H. Sun is with the Waseda Research Institute for Science and Engineering, Tokyo 169-8555, Japan and JST, PRESTO, 4-1-8 Honcho, Kawaguchi, Saitama, 332-0012, Japan (e-mail: hemingsun@aoni.waseda.jp).},
        Jiro~Katto,
        Xiaoyang~Zeng,
        and~Yibo~Fan\thanks{Y. Fan is with the State Key Laboratory of ASIC and System, Fudan University, Shanghai 200433, China (e-mail: fanyibo@fudan.edu.cn).}}

\markboth{
}
{LIU \MakeLowercase{\textit{et al.}}: Learned Video Compression with Residual Prediction and Loop Filter}
\maketitle

\begin{abstract}
In this paper, we propose a learned video codec with a residual prediction network (RP-Net) and a feature-aided loop filter (LF-Net). For the RP-Net, we exploit the residual of previous multiple frames to further eliminate the redundancy of the current frame residual. For the LF-Net, the features from residual decoding network and the motion compensation network are used to aid the reconstruction quality. To reduce the complexity, a light ResNet structure is used as the backbone for both RP-Net and LF-Net. Experimental results illustrate that we can save about 10\% BD-rate compared with previous learned video compression frameworks. Moreover, we can achieve faster coding speed due to the ResNet backbone. This project is available at https://github.com/chaoliu18/RPLVC.
\end{abstract}

\section{Introduction}
Video has accounted for approximately 80\% of internet traffic in recent years\cite{forecast2019cisco}, requiring highly efficient video codecs to reduce the cost of video data transmission.
Traditional video codecs, including H.265/HEVC\cite{h265} and H.266/VVC\cite{vvc}, are based on the classical prediction, transformation, quantization, and entropy coding frameworks to tackle the complex video coding problem.
While achieving excellent compression efficiency, they are accompanied by the following problems.
1. Each submodule relies on manual design, which makes it difficult to optimize the codec from a holistic perspective, resulting in a long iteration cycle.
2. Most of the traditional codecs are optimized for mean square error (MSE), leading to relatively poor performance in other evaluation metrics.
3. The transform module is designed with lossless bijections to concentrate the energy of the original signal, which limits the codec's ability to process and compress the data\cite{rippel2019learned}.

With the remarkable development of neural network technology, learning-based image codecs have shown great potential \cite{liu2019deep, ma2019image} to solve the problems existing in traditional coding. One of them is the famous variational autoencoders (VAEs)\cite{balle2018variational} based approach, which has achieved great performance with subsequent efforts \cite{minnen2018joint, cheng, wang2020ensemble}. Further, many learned video codecs have emerged to explore how to reduce the spatial-temporal redundancy \cite{chen2019learning, liu2020neural} in video compression. They currently contain two main types, a P-frame compression strategy with one-way reference \cite{chen2019learning, lu2019dvc, liu2020neural, lin2020m, hu2020improving} and a B-frame compression strategy with two-way reference\cite{yang2020learning, wu2018video, djelouah2019neural}.
For P-frame compression,
Chen \textit{et al.} \cite{chen2019learning} designed PMCNN with motion extension and
hybrid prediction networks to reduce spatiotemporal redundancy.
Liu \textit{et al.} \cite{liu2020neural} proposed NVC with joint spatiotemporal prior aggregation to reduce the correlations in inter frames.
Lu \textit{et al.} \cite{lu2019dvc} implemented an end-to-end deep video compression framework by using neural network modules to replace each traditional coding module. They used the optical flow to represent motion information of video and applied two VAEs to compression of the optical flow and residuals.
Lin \textit{et al.} \cite{lin2020m} employed a multi-reference frame strategy utilizing information from the reconstructed frames and the reconstructed motion vector (MV).
For the compression of B frames, Yang \textit{et al.} \cite{yang2020learning} designed an allocation strategy and recurrent enhancement to achieve hierarchical compression of video.
Wu \textit{et al.} \cite{wu2018video} designed an interpolation-based video compression network using block-level motion estimation.
Djelough \textit{et al.} \cite{djelouah2019neural} implemented an optical flow compression network that simultaneously decodes the optical flow and interpolation coefficients.
%
%
\begin{figure*}[tbp]
  \centering
  \includegraphics[width=18cm]{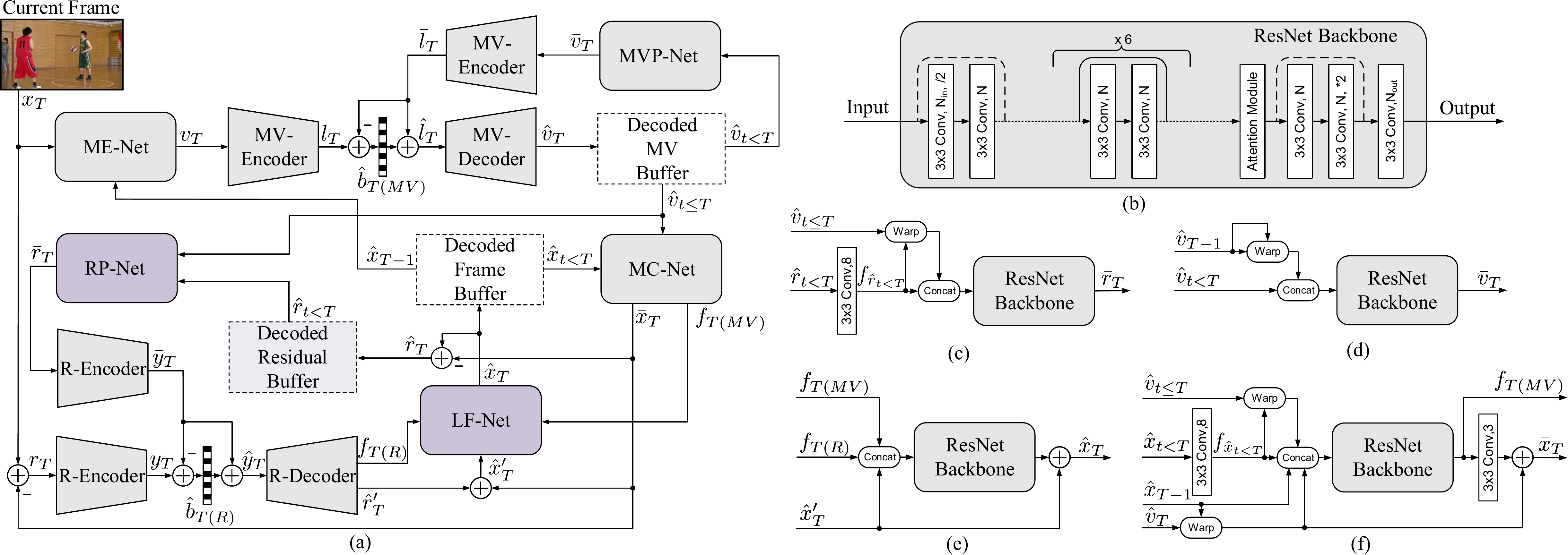}
  \caption{(a) The overall framework. (b) The ResNet backbone with attention module \cite{cheng} for submodules. (c) RP-Net, Nin=64, N=64, Nout=3. (d) MVP-Net, Nin=8, N=32, Nout=2. (e) LF-Net, Nin=195, N=128, Nout=3. (f) MC-Net, Nin=70, N=64, Nout=64. Compared to the previous work\cite{lin2020m}, we propose two new modules RP-Net and LF-Net highlighted in purple. A Decoded Residual Buffer is also added to store the decoded residuals of the previous frame. The Motion Vector Buffer stores the decoded MV of the three most recent ones, while the Decoded Residuals Buffer and Reference Frames Buffer store the reference information of four frames. ME-net is implemented by SPy-Net \cite{spynet}, while the MV and residual compression use hyperprior structures \cite{balle2018variational} with mean-scale model \cite{minnen2018joint}. The warp operation is implemented by bilinear interpolation. For the residual decoder (R-Decoder), the number of channels of its output features $f_{T(R)}$ is set to 128. The pixel-level residual $\hat{r}_{T}'$ is generated by using a convolutional layer at the tail, which behaves similarly to MC-Net.}\label{fig1}
\end{figure*}

For P-frame compression scenarios, we further pursue better coding performance based on the work MLVC of Lin \textit{et al.} \cite{lin2020m} in this paper. Compared to coding the current frame with only the reconstructed frame and MV, the residual containing rich high frequency features is sufficiently exploited in our method. By making full use of them, we effectively improve the rate-distortion performance of the proposed codec.
Also, we propose an efficient loop filtering network with exploring the feature-level information in video coding. Our specific contributions are as follows.
\begin{itemize}
  \item We propose RP-Net in this paper, which uses the reconstructed residual of the previous frames to predict the residuals of the current frame and thus reducing the temporal redundancy in the residuals.
  \item We design LF-Net that makes fuller use of the output information from the previous modules. By feeding feature-level residuals and predicted frames, the loop filter can perform higher-level texture extraction and feature learning than low-dimensional pixel-level features.
\end{itemize}

\section{Proposed Method}
\subsection{Overview of the Proposed Method}
Our architecture inherits the framework of MLVC. The difference is that we propose two new modules, RP-Net and LF-Net, to improve the performance. The workflow is as follows.
The current frame $x_T$ and the reference frame $\hat{x}_{T-1}$ are fed to the motion estimation network (ME-Net) to obtain the motion vectors $v_T$.
Before encoding it, motion prediction is first performed.
The reference MVs $\hat{v}_{t<T}$ from Decoded MV Buffer are sent to motion vector prediction network (MVP-Net) to get the predicted MV $\bar{v}_T$.
By encoding $v_T$ and $\bar{v}_T$ with the same MV encoder, the difference between the output latent code $l_T$ and $\bar{l}_T$ is quantized and sent to the entropy codec to get the first part of bits $\hat{b}_{T(MV)}$.
At the MV decoding side, the quantized difference from entropy decoder is added with $\bar{l}_T$ to recover the latent code $l_T$ at the encoding side, which is marked as $\hat{l}_T$ because of quantization noise.
Latent code $\hat{l}_T$ is then fed to the MV decoder to obtain the decoded MV $\hat{v}_T$.
We send it and the reference frames $\hat{x}_{t<T}$ to the motion compensation network (MC-Net) to get the predicted frame $\bar{x}_T$ and the corresponding feature $f_{(T)MV}$. The residuals $r_T$ between $\bar{x}_T$ and the current frame $x_T$ are sent to the residual encoder and decoder networks to get the second part of bits $\hat{b}_{T(R)}$.
Different from MLVC, we also predict the residuals by RP-Net to reduce the redundancy between residuals.
The feature-level information $f_{(T)MV}$ and $f_{(T)R}$ output from residual decoder is fed to LF-Net to obtain an enhanced reconstructed frame $\hat{x}_T$, which is cached in the buffer for the coding of the following frames. Besides, the decoded MV $\hat{v}_T$ and the difference between $\hat{x}_T$ and $\bar{x}_T$ are cached in the Decoded MV and Residual Buffer, respectively.

\subsection{Motion Prediction and Compensation}
To remove the temporal correlation of MVs, we use a MVP-Net to predict the current MV. In addition to the three decoded reference MVs, a fourth MV is created by warping the previous MV $\hat{v}_{T-1}$ itself. We concatenate these four MV and feed them to MVP-Net to obtain the predicted MV $\bar{v}_T$ as shown in Fig. \ref{fig1}(d).
\begin{equation}\label{eq:motion}
  \bar{v}_T = ResNet_{mvp}(\hat{v}_{t<T}, \text{Warp}(\hat{v}_{T-1}, \hat{v}_{T-1}))
\end{equation}
By leveraging $\bar{v}_T$, the number of required bits for compressing the MV is reduced \cite{lin2020m}.

After that, the decoded MV $\hat{v}_T$ and the reference frame $\hat{x}_{t<T}$ are fed to MC-Net to obtain the predicted frame $\bar{x}_T$ and its features $f_{T(MV)}$. We first obtain the reference frame features through a single-layer convolutional network $H_x$.
\begin{equation}\label{hmc}
  f_{\hat{x}_{t<T}} = H_x(\hat{x}_{t<T})
\end{equation}
And then we perform a warp operation on it to get the warped feature. Similarly, the previous reference frame $\hat{x}_{T-1}$ is also warped.
The reference frame $\hat{x}_{T-1}$, the warped reference frame $\text{Warp}(\hat{x}_{T-1}, \hat{v}_{T})$, the features $f_{\hat{x}_{t<T}}$ and the warped features $\text{Warp}(f_{\hat{x}_{t<T}}, \hat{v}_{t\leq T})$ are fed together into the ResNet backbone to obtain $f_{T(MV)}$ as shown in Fig. \ref{fig1}(f).
\begin{align}\label{eq:mc}
  f_{T(MV)} = ResNet_{mc}(&\hat{x}_{T-1}, \text{Warp}(\hat{x}_{T-1}, \hat{v}_{T}), \notag\\
   &  f_{\hat{x}_{t<T}}, \text{Warp}(f_{\hat{x}_{t<T}}, \hat{v}_{t\leq T}))
\end{align}
Finally, $f_{T(MV)}$ are transformed back to the pixel domain by a conversion convolutional layer $H_f$, and a straight connected warped frame is added to get the prediction frame $\bar{x}_T$.
\begin{equation}\label{eq:mc2}
  \bar{x}_T = H_f(f_{T(MV)}) + \text{Warp}(\hat{x}_{T-1}, \hat{v}_{T})
\end{equation}

\subsection{Residual Prediction Network}
Apart from the motion, the residual also plays a fundamental role in video compression and often directly reflects the amount of high frequency information in the video.
In our proposed method, the residual is defined as the difference between the prediction frame $\bar{x}_T$ from MC-Net and the original frame $x_T$.
As we known, the prediction of MC-Net is limited by the structure and parameters of the network itself. For example, the prediction may cause some blurring due to interpolation in most of the motion compensation modules.
For some complex textures appearing in different frames, the motion models may not be able to predict them, which will result in the codec always needing to encode the same residuals caused by the textures.
RP-Net is designed to reduce this redundancy by using the reference residuals $\hat{r}_{t<T}$ to predict the current residual $r_T$. The input to the RP-Net contains $\hat{r}_{t<T}$ and $\hat{v}_{t<T}$, and the output is the predicted residuals $\bar{r}_{T}$. To make full use of the information contained in the residuals, a feature extraction network $H_r$ is used to obtain the features of $\hat{r}_{t<T}$, which are then warped using $\hat{v}_{t<T}$ to obtain the warped features.
\begin{equation}\label{fen}
  f_{\hat{r}_{t<T}} = H_r(\hat{r}_{t<T})
\end{equation}
The network $H_r$ contains only one convolutional layer to avoid excessive complexity. The features $f_{\hat{r}_{t<T}}$ and the warped features $\text{Warp}{(f_{\hat{r}_{t<T}}, \hat{v}_{t<T})}$ are concatenated and sent to the ResNet backbone to obtain $\bar{r}_{T}$.
\begin{equation}\label{rpnet}
  \bar{r}_{T}= ResNet_{rp}(f_{\hat{r}_{t<T}}, \text{Warp}{(f_{\hat{r}_{t<T}}, \hat{v}_{t<T})})
\end{equation}
Similar to the effect of MVP-Net in MV compression, the number of required bits for coding the residual is also reduced with RP-Net.

\subsection{Feature-aided Loop Filter}
After completing the motion-related and residual-related part of the coding, the obtained prediction $\bar{x}_T$ and residual $\hat{r}_T'$ is finally fed to the LF-Net to reduce the artifacts.
The loop filter network itself does not use extra bits to represent information as a compression network does, so how to make the most of the information obtained from the previous stage is key to designing the filter.
To address this, our design utilizes feature-level rather than pixel-level information. This is adopted for the following two points.
1. In traditional coding, the image is only filtered at the pixel level, which typically contains only 3 layers of pixel information: RGB or YUV. If this strategy is used in a neural network codec, the residual and the predicted frame are also need to be converted to the pixel level before being fed into the filter, which results in a large number of features learned in the precursor networks being discarded. To avoid this, we also extract the features from the middle of the predecessor network as the input to the filter to make the whole framework run more efficiently.
2. Inputting reference information into the filter network can lead to duplication and redundancy because the same extraction of low-level textures has been performed in the predecessor networks. Therefore, the reference information in our model, including MV, residuals and frames, is only fed to their corresponding prediction modules, and the features output from previous modules are used as the input to the loop filter.
More specifically, we extract the features $f_{T(R)}$ and ${f_{T(MV)}}$ from residual decoder and MC-Net as the input to the loop filter.
\begin{equation}\label{lfnet}
  \hat{x}_{T}= ResNet_{lf}(f_{T(MV)}, f_{T(R)}, \hat{x}_{T}')
\end{equation}
In addition, the sum $\hat{x}_T'$ of the predicted frames $\bar{x}_T$ and the pixel-level residuals $\hat{r}_T'$ is also used as input to the filter, which acts as a shortcut to help back-propagate the gradient during training.
\begin{equation}\label{lfnet2}
  \hat{x}_{T}' = \bar{x}_T + \hat{r}_T'
\end{equation}

\begin{figure*}[t]
\centering
    \subfigure[]{
    \begin{minipage}[t]{0.33\linewidth}
    \centering
    \includegraphics[scale=0.37]{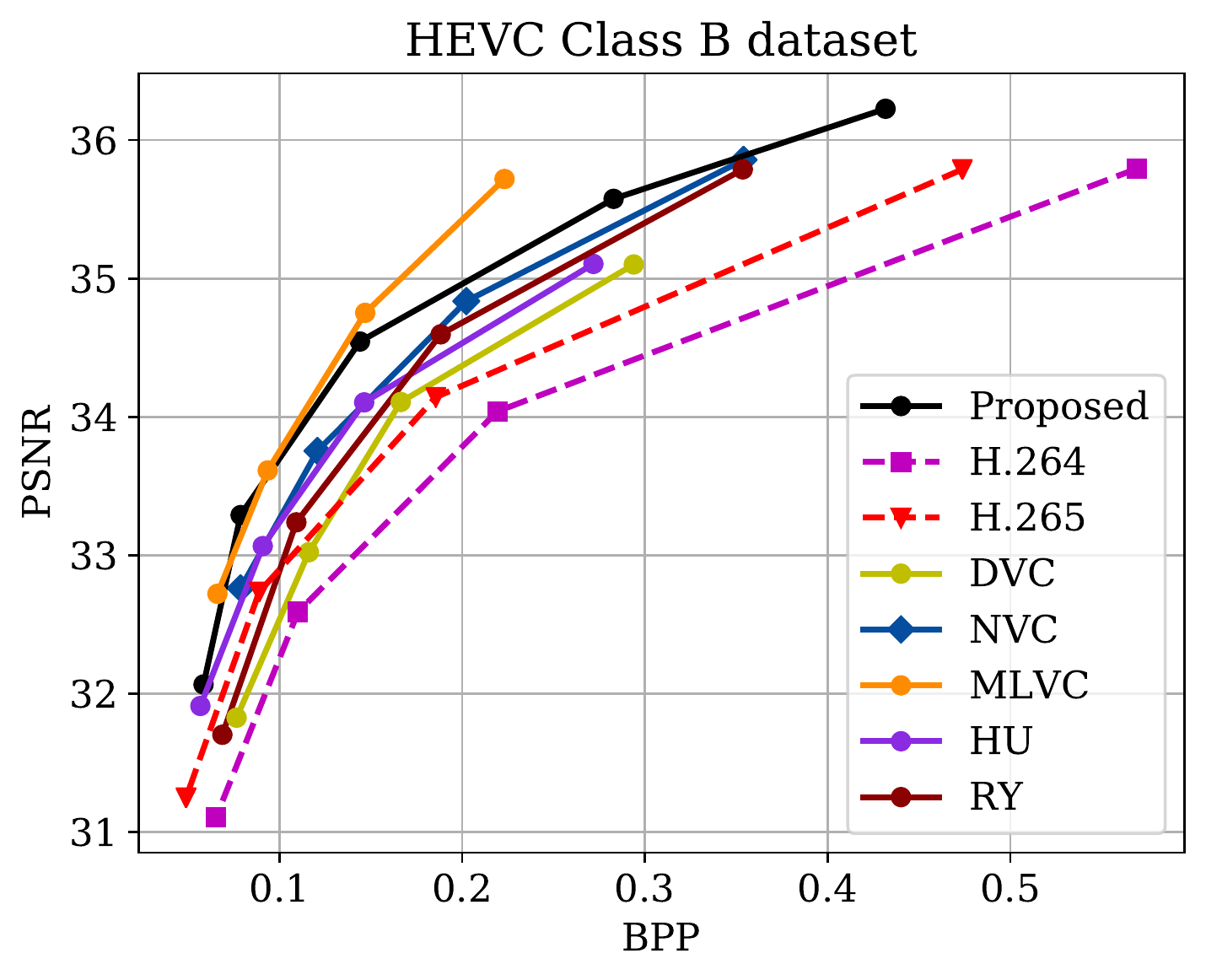}
    \end{minipage}%
    }%
    \subfigure[]{
    \begin{minipage}[t]{0.33\linewidth}
    \centering
    \includegraphics[scale=0.37]{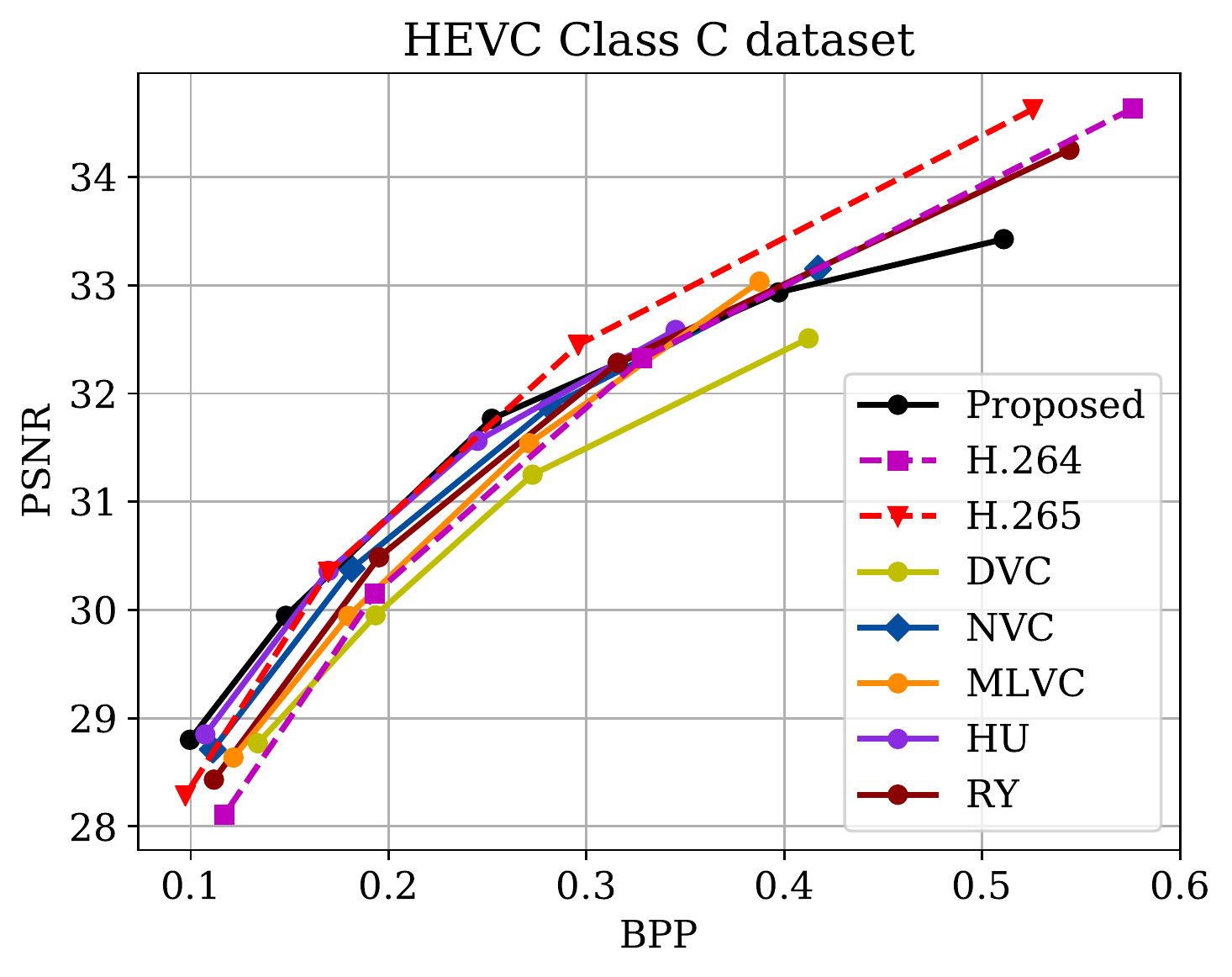}
    \end{minipage}%
    }%
    \subfigure[]{
    \begin{minipage}[t]{0.33\linewidth}
    \centering
    \includegraphics[scale=0.37]{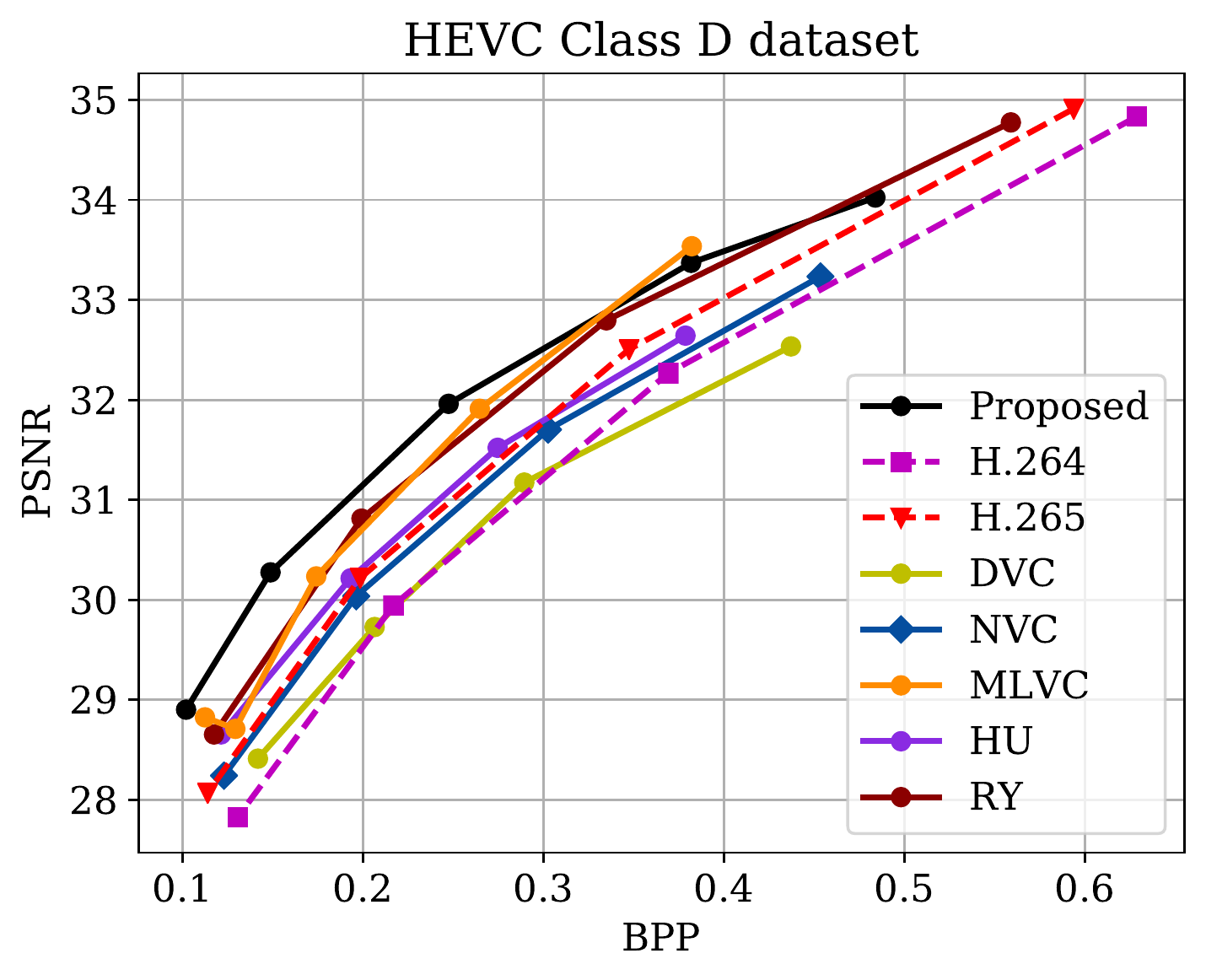}
    \end{minipage}%
    }%

    \subfigure[]{
    \begin{minipage}[t]{0.33\linewidth}
    \centering
    \includegraphics[scale=0.37]{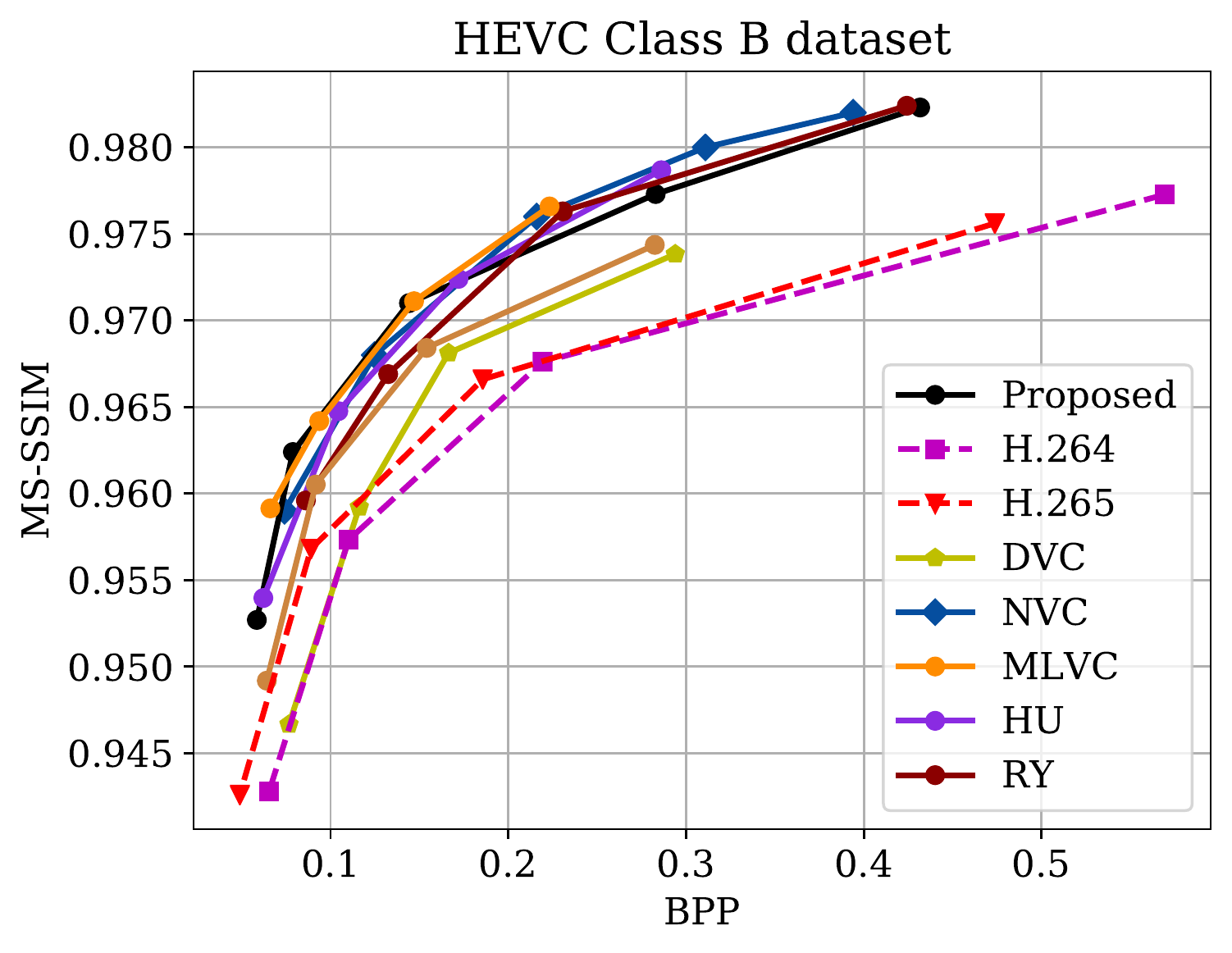}
    \end{minipage}%
    }%
    \subfigure[]{
    \begin{minipage}[t]{0.33\linewidth}
    \centering
    \includegraphics[scale=0.37]{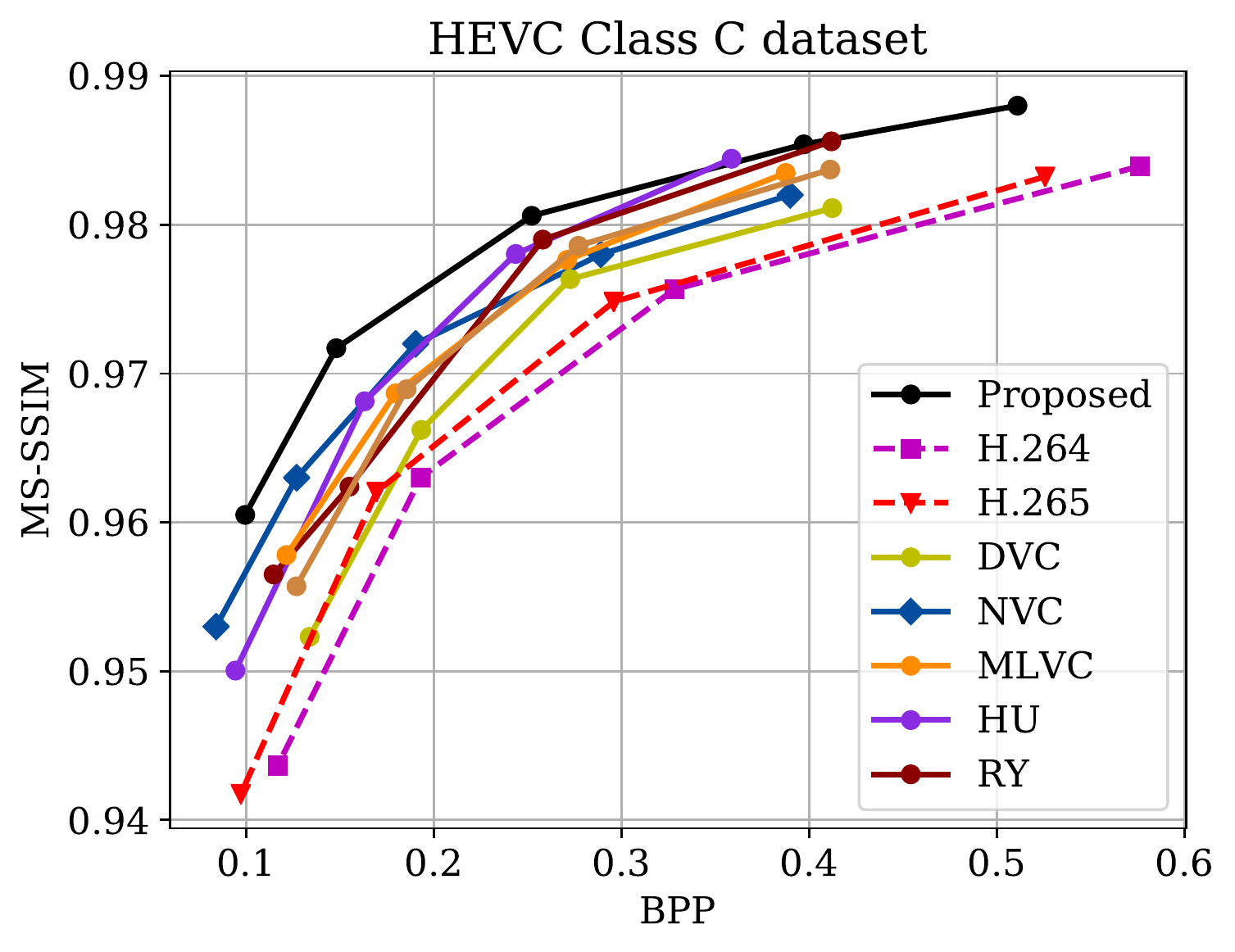}
    \end{minipage}%
    }%
    \subfigure[]{
    \begin{minipage}[t]{0.33\linewidth}
    \centering
    \includegraphics[scale=0.37]{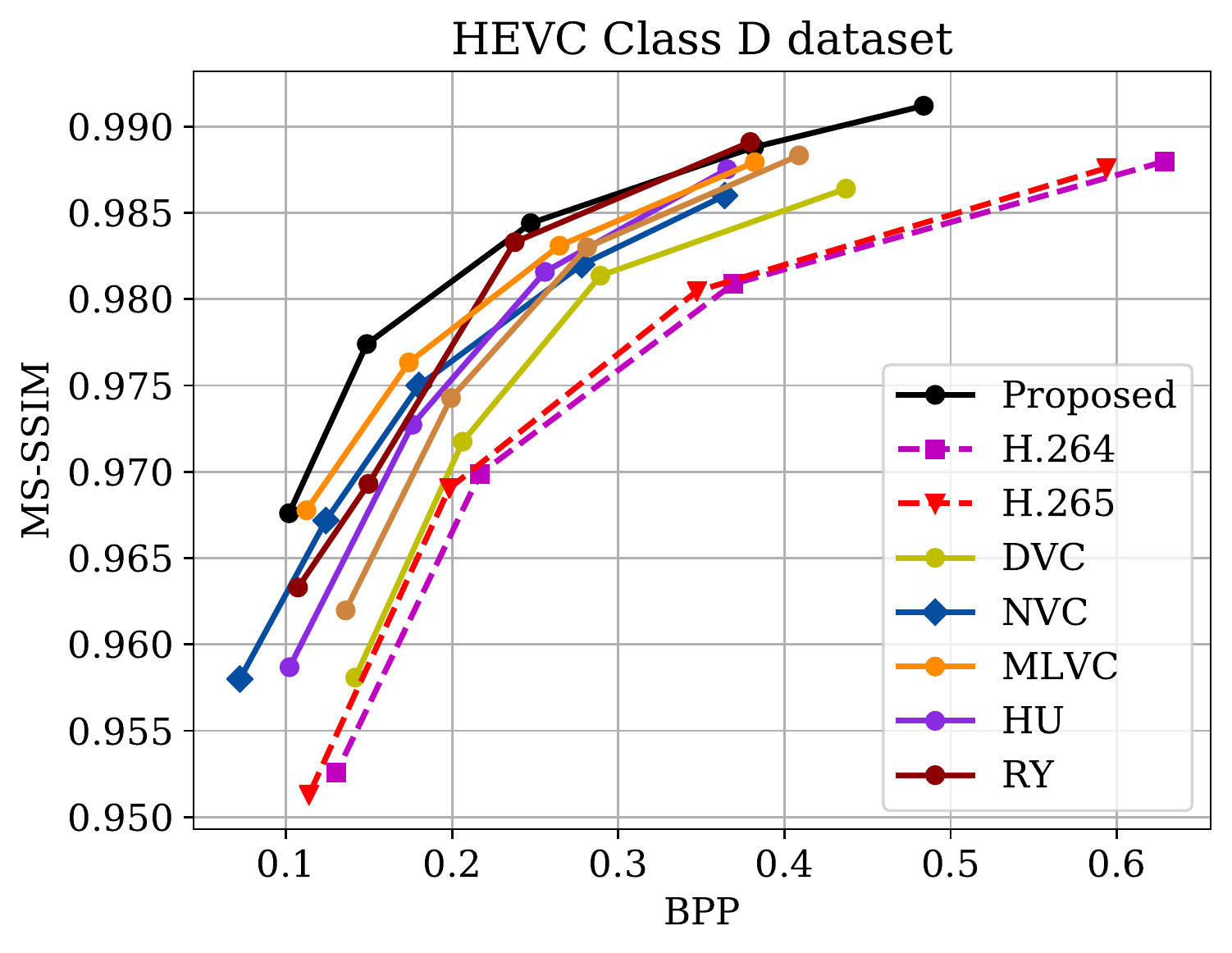}
    \end{minipage}%
    }%
\caption{Overall performance. The comparison is conducted with NVC\cite{liu2020neural}, DVC\cite{lu2019dvc}, MLVC\cite{lin2020m}, HU\cite{hu2020improving}, and RY\cite{yang2020learning}. The results of H.264 and H.265 were obtained by ffmpeg with medium preset \cite{pvc}. The results of the work \cite{lu2019dvc, hu2020improving, yang2020learning, lin2020m, liu2020neural} are cited in their original papers or their open source projects.}\label{objective}
\end{figure*}

\section{Experiment}
\subsection{Experimental Setting}
\textbf{Training.} We used the dataset Vimeo90K \cite{xue2019video} and MIT-2 \cite{monfort2019multimoments} to train our proposed models. At the beginning of the training, only a single P-frame was used. The number of P-frames used in training was gradually increased as the number of training iterations increases. Specifically, we updated the training clip by one P-frame after every 5e4 iterations and overall approximately 1e6 iterations were trained using the Adam optimizer \cite{kingma2015adam}. The batch size was set to 4. With classical rate-distortion loss, we trained 5 models, first using PSNR with $\lambda$ \{512, 1024, 2048, 4096, 6144\} as quality metrics for optimization \cite{begaint2020compressai} and then fine-tuning them using MS-SSIM index. The 1-st frames of the video were encoded with BPG \cite{bpg} and repeated four times as an initial buffer of reference frames. The initial reference motion vector and residuals in the buffer were set as zero tensors. Each subsequent P-frame was encoded with the reference buffer and the buffer was updated with the coding process.

\textbf{Evaluation.} To demonstrate the robustness of our model, the same datasets and test condition, including HEVC test datasets (classes B, C, D, and E)\cite{bossen2013common} and UVG datasets \cite{uvg}, with the reference work \cite{lu2019dvc} were used. These videos were coded with IPPP coding structure.
The average bits per pixel (BPP) and PSNR/MS-SSIM were utilized as metrics to evaluate the bit rate and reconstructed quality, respectively. Based on them, we plotted RD curves and calculated BD-rate \cite{bjontegaard2001calculation} for qualitative and quantitative comparisons. All of the previously mentioned experiments were conducted on the platform with Intel Xeon Gold 6230 CPU @ 2.10GHz and Nvidia RTX 2080 Ti GPU.

\begin{table*}[tbp]
  \centering
  \caption{BD-rate Reduction Comparison with NVC \cite{liu2020neural} and MLVC \cite{lin2020m} (Anchor: H.265)}
  \setlength{\tabcolsep}{1mm}{
    \begin{tabular}{c|r|r|r|r|r|r}
    \hline
    \multirow{2}[4]{*}{Videos} & \multicolumn{3}{c|}{PSNR} & \multicolumn{3}{c}{MS-SSIM} \bigstrut\\
\cline{2-7}          & \multicolumn{1}{c|}{NVC} & \multicolumn{1}{c|}{MLVC} & \multicolumn{1}{c|}{Proposed} & \multicolumn{1}{c|}{NVC} & \multicolumn{1}{c|}{MLVC} & \multicolumn{1}{c}{Proposed} \bigstrut\\
    \hline
    ClassB & -25.81\% & -43.74\% & -32.16\% & -40.97\% & -40.11\% & -40.45\% \bigstrut\\
    \hline
    ClassC & 8.10\% & 11.67\% & 0.33\% & -27.31\% & -12.49\% & -40.70\% \bigstrut\\
    \hline
    ClassD & 4.89\% & -12.15\% & -21.88\% & -33.58\% & -28.82\% & -47.45\% \bigstrut\\
    \hline
    ClassE & -18.98\% & -15.43\% & -35.20\% & -42.08\% & -22.27\% & -34.49\% \bigstrut\\
    \hline
    UVG   & -10.64\% & -17.56\% & -22.50\% & -38.83\% & -22.98\% & -27.23\% \bigstrut\\
    \hline
    Average & -8.49\% & -15.44\% & \textbf{-22.28\%} & -36.55\% & -25.33\% & \textbf{-38.07\%} \bigstrut\\
    \hline
    \end{tabular}%
    }
  \label{bdrate}%
\end{table*}%

\subsection{Performance Evaluation}\label{pe}
\textbf{Rate-distortion Performance.}
As shown in Fig. \ref{objective}, we compare our model with traditional codecs (H.264 and H.265) and  previous learned works \cite{lu2019dvc, hu2020improving, yang2020learning, lin2020m, liu2020neural}. Compared with them, our model achieves a favorable performance under HEVC Class B, C, and D datasets. The BD-rate results with H.265 anchor is shown in Table \ref{bdrate}. Compared to H.265, our model achieves about 22.28\% and 38.07\% BD-rate gains in PSNR and MS-SSIM metrics, respectively. Our model also achieves a gain of more than 10\% than NVC in PSNR metric. Compared with MLVC, the proposed method achieves about 10\% BD-rate reduction. It is worth mentioning that we used the same model to obtain the test results of PSNR and MS-SSIM metrics. It is believed that the proposed method can achieve better performance if it is optimized with different metrics separately.

\begin{figure}
  \centering
  \includegraphics[width=5.5cm]{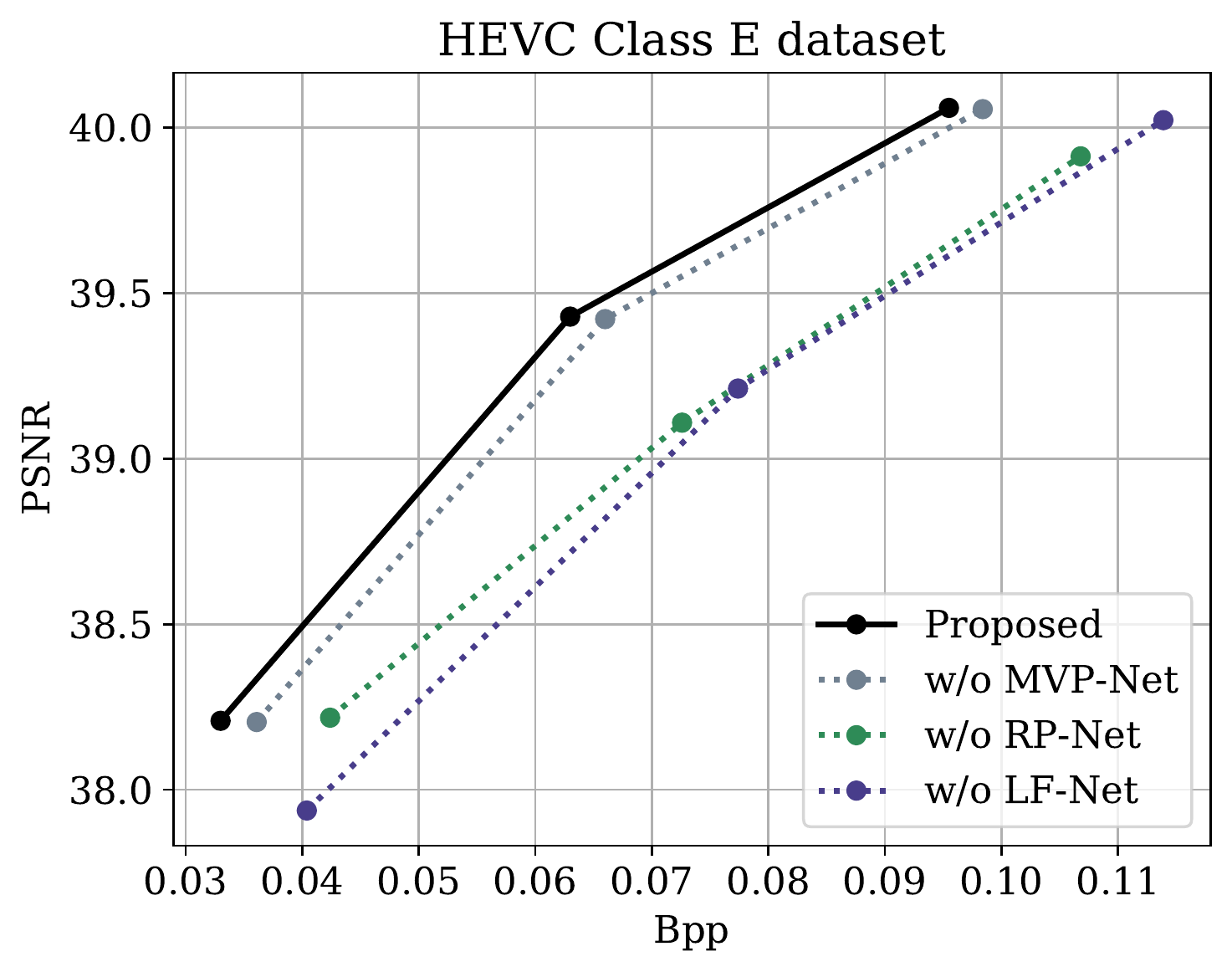}
  \caption{Ablation study. Three ablation experiments were performed on the proposed method with $\lambda$ in \{2048, 4096, 6144\}.}\label{ablation}
\end{figure}

\begin{figure}
  \centering
  \includegraphics[width=8.8cm]{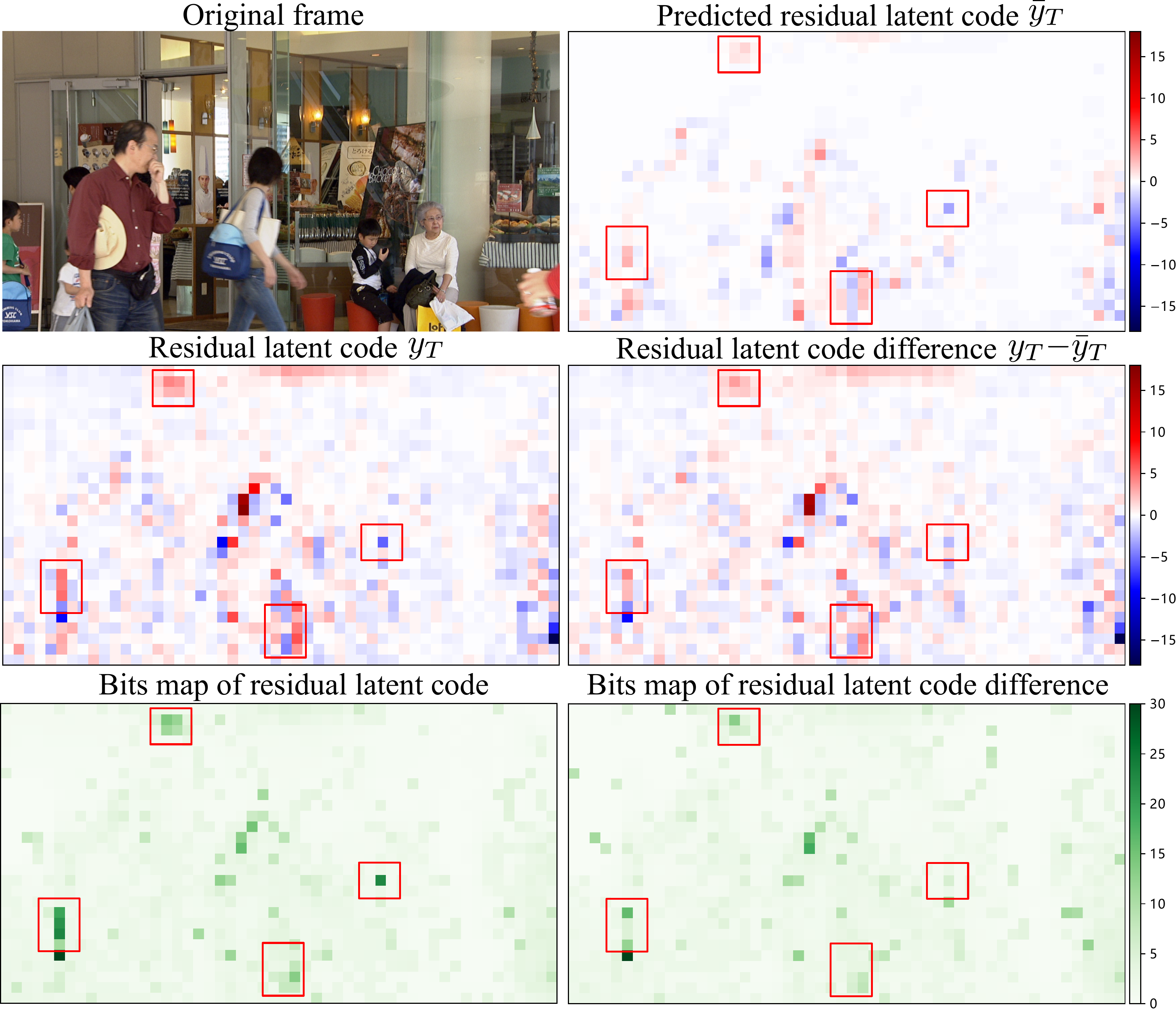}
  \caption{Visualization comparison of compressing the BQMall sequence using RP-Net with $\lambda = 4096$. With RP-Net, the absolute values in the red boxes of the latent code and the corresponding bits map are reduced, implying that the number of required bits to represent the latent code is reduced.}\label{ablationRP}
\end{figure}

\textbf{Ablation Study.} 
As shown in Fig. \ref{ablation}, we remove MVP-Net, RP-Net, and LF-Net from the proposed model and performed comparisons. It can be found that the performance of the model deteriorates to some extent after removing these modules, which demonstrates the necessity of these modules.
For RP-Net, we visualize the intermediate tensors as shown in Fig. \ref{ablationRP}.
It can be seen that the predicted residual latent code $\bar{y}_T$ and the residual latent code $y_T$ have some similarities in some pixels. By making a difference between the two, the absolute values of the latent code are reduced, implying the number of bits required to represent the $y_T$ is effectively reduced. 

\begin{figure}
  \centering
  \includegraphics[width=8.5cm]{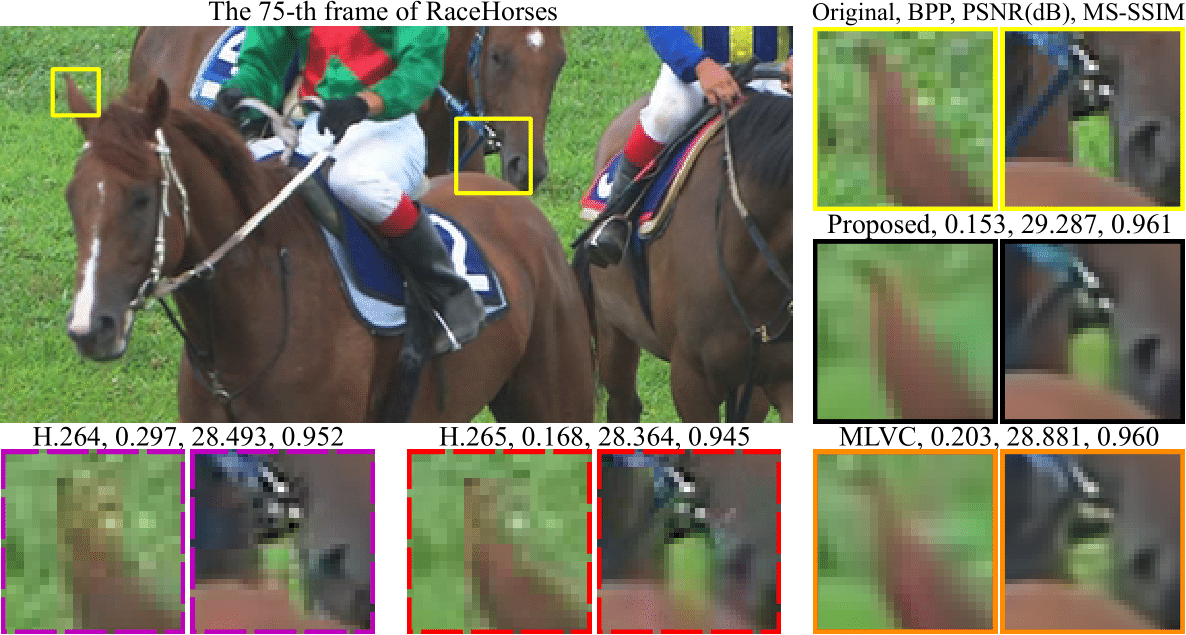}
  \caption{Visualization comparison of compressing the RaceHorses sequence with $\lambda = 512$.}\label{subjective}
\end{figure}

\begin{table*}[tbp]
  \centering
  \caption{Comparison of Complexity and Coding Speed with MLVC \cite{lin2020m}}
    \setlength{\tabcolsep}{1mm}{
    \begin{tabular}{c|c|c|c|c|c|c|c}
    \hline
    \multirow{2}[4]{*}{Methods} & \multicolumn{2}{c|}{Complexity} & \multicolumn{5}{c}{Speed (fps)} \bigstrut\\
\cline{2-8}          & Params & FLOPs & Type  & ClassB & ClassC & ClassD & ClassE \bigstrut\\
    \hline
    MLVC  & 167.3M & 3.0T  & Enc.   & 0.24  & 1.28  & 3.46  & 0.48 \bigstrut\\
    \hline
    \multirow{2}[4]{*}{Proposed } & \multirow{2}[4]{*}{27.1M} & \multirow{2}[4]{*}{2.7T} & Enc.   & 0.35  & 1.67  & 5.46  & 0.78 \bigstrut\\
\cline{4-8}          &       &       & Dec.   & 0.62  & 3.24  & 9.09  & 1.51 \bigstrut\\
    \hline
    \end{tabular}%
    }
  \label{complexity}%
\end{table*}%

\textbf{Model Complexity and Coding Speed.}
We have made two refinements to the proposed model to ensure it has low complexity. First, the basic ResNet backbone in our method uses computations in the downsampling domain, reducing the amount of computation by about 75\%. Second, we removed the residual refine and MV refine networks in MLVC to make the proposed model more compact, so our model achieves a lower complexity.
As shown in Table \ref{complexity}, our model has fewer model parameters, lower FLOPs (720p) and relatively faster coding speed compare with MLVC.


\textbf{Subjective Quality Evaluation.} As shown in Fig. \ref{subjective}, we perform a subjective comparison of the proposed method with H.264, H.265, and MLVC. It can be found that our model retains the texture information of the horse edge and rope with significantly fewer bits (BPP), while both H.264 and H.265 show some artifacts and distortion and the learned method MLVC appears somewhat blurred.

\section{Conclusion}
We propose two novel modules for the learned video codec in this work: a residual prediction module and a feature-aided loop filter. The corresponding ablation experiments validate the effectiveness of these modules. We also compare our proposed method with many previous works, including both traditional and learned methods. In particular, our model achieves a BD-rate gain of about 30\% and 10\% compared to H.265 and MLVC, respectively. For the complexity compared with the learned method MLVC, our model achieves faster coding speed and lower FLOPs with fewer parameters.


\bibliographystyle{IEEEtran}
\bibliography{reference}
\end{document}